\documentclass[12pt]{article}
\usepackage{a4}
\usepackage{graphicx}
\usepackage{amsfonts}
\usepackage{amscd}
\usepackage{latexsym}
\usepackage{epsf}
\usepackage{amsmath}
\numberwithin{equation}{section}

\newcommand{\be}{\begin{equation}}
\newcommand{\ee}{\end{equation}}
\newcommand{\eom}{equation of motion }
\newcommand{\bea}{\begin{eqnarray}}
\newcommand{\eea}{\end{eqnarray}}

\newcommand{\eqq}{equation }
\newcommand{\eqqs}{equations }

\newcommand{\fr}{\frac}
\newcommand{\pd}{\partial}

\newcommand{\half}{\frac{1}{2}}

\renewcommand{\psi}{\varphi}
\begin{document}

\title{Local Energy Velocity of Classical Fields\\ \bf\small Language to be improved
} 
\author{
{\sc I.~V.~Drozdov }\thanks{e-mail: drosdow@uni-koblenz.de} \\
{\small and}\\
A.~A.~Stahlhofen \thanks{e-mail: alfons@uni-koblenz.de}\\
\small  University of Koblenz, Institut f\"ur Naturwissenschaften\\ 
\small  Abteilung Physik\\
\small  Universit\"atsstr.1, D-56080 Koblenz, Germany}
\maketitle
 
\begin{abstract}

It is proposed to apply a recently developed concept of local wave velocities to the 
dynamical field characteristics, especially for the canonical field
energy density.  
It is shown that local energy velocities can be derived from the 
  lagrangian directly.   
 The local velocities of zero- and first- order for energy propagation has been obtained 
for special cases of scalar and vector fields.  
 Some important special cases of these results are discussed.
\end{abstract}

\section{Introduction}
 
  The terms of "propagation" and "velocity"
  supposed to characterize a wave propagation
   can in case of classical fields
be treated as a propagation of a local attribute, as shown recently
This approach provides a spectrum of local phase velocities.  
 
 These local velocities, being itself local dynamical fields, stage an 
 alternative representation of the original physical field and their
  behaviour under several important local symmetries is the subject 
  of investigation for physical applications. For instance, the two- and three-
  dimensional generalizations of zero- and first-order velocities to co-variant and
  contra-variant vectors respectively are established \cite{dimension}
 
  Especially, it has been shown \cite{WV}, that the zero-order velocity is Lorentz
  covariant in sense of the relativistic addition of velocities. This property preserves
  the local relativistic causality on the zero-order level under change of frame.  
   For local velocities of higher orders the local Lorentz covariance does not hold and the
   subluminal restriction in arbitrary frame is not assured. Moreover, simplest counter-examples
   have been demonstrated \cite{WV, dielectric} where the first order propagation velocity may be not
   restricted by $c$ for the global propagation between two points.   
     
    In the framework of discussion on a relativistic causality and its violation by
  several local velocities overcoming the fundamental constant $c$, the following remarks are in order.
  
   The relativistic causality still remains kept, as long as the attribute propagating
  is only measured in two differnt space-time points and the propagation means no 
  real transmission of "some physical feature" under consideration from the one point to
the other one. Superluminal phase and group velocities, for instance, are well known. 
  
  Analogous, the propagation velocity of zero order, i.e. propagation of the attribute itself,
  may approach unbounded values overcoming the light velocity, without to controverse the
 relativistic causality immediately.   
  
  On the contrary, one should proceed carefully with velocities of higher order. Since the propagation
  of the attribute like a derivative can be treated as one of a state alternation in a physical system (i.e. by an interaction
   process), the propagation might be interpreted as an interaction transmission or information transmission respectively.
   Therefore, this velocity may concern with a real physical phenomenon and deserves a consistent interpretation.

\section{ Local energy velocity, force velocity etc.}

  The subject of "energy velocity" being dealth with in the optics, is usually interpreted in context of continuity \eqq
  containing only the global energy density of the all matter in the space and its current flow density respectively.
   An assignment of any velocity to this current for an arbitray configured field is therefore an ambiguous
 non-local procedure.
   
  Instead of that, the energy state of a certain field of the field system can be observed, whereat   
the corresponding energy density $T^{00}$ (zero component of Noether's conserved current) can be considered as a scalar field.
  Therefore, the formalism of local velocities \cite{WV} can be applied here.
  
 Basically, for any scalar field $\psi(x,t)$ the velocity for a corresponding local attribute propagation is defined
 as the N-th order phase velocity (N-PV): 
\bea
& v_{(0)}(x,t):=\fr{dx^f}{dt}= -\fr{ \dot{\Psi} }{ \Psi' }& \ \ \mbox{ 0-PV (field velocity) }\\
& v_{(I)}(x,t):= \fr{dx^p}{dt}= -\fr{ \dot{\Psi}' }{ \Psi'' }& \ \ \mbox{ 1-PV (peak velocity) } \\
& v_{(II)}(x,t):= \fr{dx^{TP}}{dt}= -\fr{ \dot{\Psi}'' }{ \Psi''' }& \ \ \ \mbox{ 2-PV (turning point velocity)}
\eea
 and so on \cite{WV}.
 
In the approach of the field theory, the state of a physical object is decribed by the field $\Psi(x,t)$, 
treated, for instance, as a density of probability distributed in space-time. 
         
 Let us start the discussion by briefly recalling the local concept.
  The observation of a time evolution of the field state in accordance with the \eom
  provides for a fixed field value $\Psi_0$ the propagation velocity $v_{(0)}$ of zero order (0-PV).

  A certain field state is descibed by the corresponding distribution of energy density $T^{00}$; 
  the fixed value ${\cal E}$ of it propagates with the 0-PV $w_{(0)}$ 
     In this sense it deals entirely with an alternative -in this case energetic- representation of the field
   $\Psi$ by the filed ${\cal E}$. Nevertheless it does not mean an equivalence of both approaches,
    especially of two propagation velocities $v_{(0)}$ and $w_{(0)}$.
   
    In fact, the possible discrepancy is due to the structure of lagrangian for the field $\Psi$ and the 
    velocities  $v_{(0)}$ and $w_{(0)}$ are in general never equivalent.
  This fact affords no contradictions to basical statements of physics, e.g.energy conseravtion,
 since these velocities are being interpreted properly as local features. 
      Namely, an alternation of the field $\Psi$ in the point $x_0$ at the time $t_0$ implies an alternation of
  the corresponding energy density.       
      In the next moment $t_1$ the same field value $\Psi$ will be recovered in a neighbouring point $x_1$ because of the
    field "re-configuration" due to the natural field dynamics (\eom).     
	Since the variation of the energy state involves also an energy for this re-configuration,
	 the point $x_2$ where the initial local energy density is to recover, will not be the same as $x_1$.	         
  In other words, the energy density includes a kinetic contribution of the dynamical re-shaping.
   The zero-order energy velocity should involve therefor at least the first order field velocity
	together with the zero-order one.	
    
   The question, how local energy velocities are related to the local field velocities, is discussed
    in following sections of the present work.   
    A formal analysis of the lagrangian procedure is associated with the form and special properties 
    of lagrangian (e.g. symmetries) and produces no general predictions. 
    
    For instance, for the lagrangian density under consideration being an arbitrary function of the field $\Psi$
    and its first derivatives, so that
    \be {\cal L}=  {\cal L}( \Psi, \dot{\Psi}, \Psi', \Psi_k ),  \ee  
   (where the dot denotes the derivative with respect to time, the prime - with respect to the chosen direction $x$, 
    and the index $k$- with respect to other directions $\{y,z\}$), the energy density of $\Psi$ reads \cite{ryder} 
    \be 
    {\cal E}=T^{00}=\dot{\Psi} \fr{ \pd{\cal L} }{ \pd \dot{\Psi}  } - {\cal L}
    \ee
    and one obtains for the zero order energy velocity
    \be
    w_{(0)} =-\fr{
    \fr{ \pd^2{\cal L} }{ \pd\dot{\Psi}\pd\Psi  } \dot{\Psi}^2
+     \fr{ \pd^2{\cal L} }{ \pd\dot{\Psi}^2  } \ddot{\Psi}\dot{\Psi}
 +    \fr{ \pd^2{\cal L} }{ \pd\dot{\Psi}\pd\Psi_k  } \dot{\Psi}_k\dot{\Psi}
  +  \fr{ \pd^2{\cal L} }{ \pd\dot{\Psi}\pd\Psi'   }\dot{\Psi}'\dot{\Psi} 
 -   \fr{ \pd{\cal L} }{ \pd \Psi  } \dot{\Psi}
  -  \fr{ \pd{\cal L} }{ \pd \Psi_k} \dot{\Psi}_k
   - \fr{ \pd{\cal L} }{ \pd \Psi' } \dot{\Psi}'
    }{
    \fr{ \pd^2{\cal L} }{ \pd\dot{\Psi}\pd\Psi  } \dot{\Psi}\Psi'
    + \fr{ \pd^2{\cal L} }{ \pd\dot{\Psi}^2  } \dot{\Psi}\dot{\Psi}'
     +\fr{ \pd^2{\cal L} }{ \pd\dot{\Psi}\pd\Psi_k  } \dot{\Psi}\Psi'_k
    +\fr{ \pd^2{\cal L} }{ \pd\dot{\Psi}\pd\Psi'   } \dot{\Psi}\Psi''
    -\fr{ \pd{\cal L} }{ \pd \Psi  } \Psi'
    -\fr{ \pd{\cal L} }{ \pd \Psi_k} \Psi'_k
    -\fr{ \pd{\cal L} }{ \pd \Psi' } \Psi''
    } 
    \ee
      
 Especially, in view of a discussion on the relativistic causality, is it sufficient to assure 
  whether the local energy velocity could not overcome the field velocity itself or moreover 
  exceed the light velocity in vacuum.
  
 In terms introducted above the condition of relativistic causality of energy for zero order can
 be folmulated e.g. as follows:
 \be 
 \dot{\Psi}^2 \pd_m\left[ \fr{ \pd{\cal L} }{ \dot{\Psi} \pd \Psi_m} \right] \le \dot{\Psi}
   \left(\fr{ \pd{\cal L} }{ \pd\dot{\Psi} }  \right)' -\fr{ \pd{\cal L} }{ \pd \Psi  } \Psi'
    -\fr{ \pd{\cal L} }{ \pd \Psi_k} \Psi'_k
    -\fr{ \pd{\cal L} }{ \pd \Psi' } \Psi''
 \ee
 
On the one hand, several physical requirements like these mentioned above
 arrange for an interpretation of energy velocity for a given lagrangian.
  
On the other hand, they impose certain restriction on the construction of lagrangians 
describing an admissible "non-exotic" matter.   

 We proceed with local velocities of higher order for energy density in order
   to give an interpretation of the subject of interaction velocity, as outlined in the introduction. 
   
   Suppose, the field $\Psi$ expires at the moment $t_0$ a small perturbation: 
   for example, the value $\Psi_0$ changes in the point $x_0$ stepwise to the $\Psi_1$.
    According to the \eqq of motion
   ("natural dynamics") this steplike change will be observed in the point $x_1$ at the moment $t_1$. 
    A pure distributive field shape is not realisable physically and looks in fact as a kink. The velocity of the propagating 
    kink can be represented by the turning point velocity 2-PV. The perturbation of the field means also the alternation
     of energy density in this point and leads to the result, the energy density in the point $x_1$ at later moment will change as well.
      A simplest signal of changed shape is for instance the first derivative of the shape.
       In this sense, the 1-PV of energy density can be interpreted as a propagation of interaction.
        Here is it not necessary to indentify this phenomena with an "energy transmission" or a "transmission" at all. 
	  The energy density descends local in one point and as a result it ascends in another one, as well as a field value
	   $\Psi$. The subject of "transport" is irrelevant for this phenomena, since these values are not labelled to identify.
In other words, one cannot say, the energy of field density ascending in the point $x_1$ is "the same" one leaving the point $x_0$.     
    The energy density of a single field $\Psi$ is also not subjected to the conservation in form of some continuity \eqq, 
    since the lagrangian may involve any interactions with other fields, but only the total energy of the system of fields
     remains conserved.    
   The formal analysis, discussed above, will be demonstrated now on simplest and well known field-theoretical models.
  
  \subsection{Scalar field}
  
  For a 1-dimensional field shape $\psi=\psi(x,t)$ of a massive scalar field
 with the lagrangian density
\be
{\cal L}=\half \pd_\mu\psi \pd^\mu\psi-\half m^2 \psi^2 - V(\psi)=
\half \{ \dot{\psi}^2-\psi'^2-m^2\psi^2 \}- V(\psi), 
\ee
$(c=\hbar=1)$  the corresponding energy density reads:
\be {\cal E}= T^{00} (x,t)=\half \{  \dot{\psi}^2+\psi'^2+m^2\psi^2 \} + V(\psi).\ee

The zero order energy propagation velocity $w_{0}$  is defined as
\be  w_{(0)}= -  \dot{ {\cal E} }/  {\cal E}'.
\label{zero_def}
\ee

 Additionally, it holds for the filed $\psi$ the \eqq of motion (Klein-Gordon wave \eqq)
\be \ddot{\psi}+m^2\psi + \fr{\pd V}{\pd\psi}  =\psi''.
\label{EOM}
\ee 
This \eqq, derived once again with respect to $x$ provides the relation:
\be   \ddot{\psi}'+m^2\psi' + \fr{\pd^2 V}{\pd\psi^2} \psi'  =\psi'''
\label{EOM_deriv}
 \ee  

Now, a direct application of the definition (\ref{zero_def}) of zero-order phase velocity (0-PV) to the energy density $T^{00}(x,t)$ 
gives:

\be
w_{(0)}=  - \fr{v_{(0)}+v_{(I)}}{1+ v_{(0)}v_{(I)} + \fr{1}{\psi''} \left[ m^2\psi + \fr{\pd V}{\pd\psi} \right] }= 
\fr{v_{(0)}+v_{(I)} +2m^2\psi}{v_{(0)}v_{(I)} + \fr{ \ddot{\psi} }{\psi''}  },
\ee
where $ v_{(0)}, v_{(I)} $ are the 0- and 1-PV's of the field shape respectively. The \eom (\ref{EOM}) is in this realtion encountered.

 For massles fields with a constant potential it follows, the zero-order energy velocity is a relativstic
addition of zero-(field) and first order (peak) velocities of the field.
 
Proceeding in a similar way with the first order energy velocity and using the relations (\ref{EOM},\ref{EOM_deriv}), one obtains: 
\be
w_{(I)}= \fr{v_{(0)}+v_{(II)} +2 C v_{(I)} }{C v_{(I)}^2 + v_{(0)}v_{(II)} + 
\fr{C}{\psi''} \left[ m^2\psi + \fr{\pd V}{\pd\psi} +\psi''\right] +\fr{1}{\psi'''} \left[ m^2\psi' + \fr{\pd^2
V}{\pd\psi^2} \psi'+\psi''' \right] }
\ee
with the notation \be C:= \fr{\psi''^2}{\psi'''\psi'}  \ee 
for the dimensionsless factor $C$, dependent on the "shape curvature" of the field. It follows from the relation of $w_{(I)}$,
  for the turning point (TP) as a field shape attribute being traced 
 $( \psi''=0 )$ the curvature factor disappears and the expression becomes similar to that for the   $w_{(0)}$:
 
 \be
 w^{TP}_{(I)} = \fr{ v_{(0)}+v_{(II)} } { v_{(0)}v_{(II)}  +\fr{1}{\psi'''} \left[ m^2\psi' + \fr{\pd^2
V}{\pd\psi^2} \psi'+\psi''' \right] }
 \ee

 Hence we can see again, for a massless scalar field $\psi$ with the potential $V$ linear with respect to the field,
 the first order local velocity of the perturbation of the energy density is a relativstic addition of zero-(field velocity)
 and second order (turning point) velocities. 
   This velocity can be interpreted as a propagation of a step-like (or "dicontinuity point") singal in the scalar field.

 \subsection{Massless vector field (electrodynamics)}
    
Starting with the Fermi lagrangian   
\be
{\cal L}=-\fr{1}{4} F_{\mu\nu}^2, \ \ \ \mbox{ with }   \ \  F_{\mu\nu}\equiv\pd_\mu A_\nu-\pd_\nu A_\mu
\ee
 one obtans the source-free Maxwell \eqq
 \be
    \pd_\nu F^{\mu\nu}=0
 \ee 
and the energy-momentum tensor \cite{LL}
\be
T^\mu_\nu=\left(  F^{\mu\rho}F_{\rho\nu} - \fr{1}{4} \delta^\mu_\nu F^2  \right),
\ee
that means for the energy density in terms of usual physical fields: 
\be
 {\cal E}= T^0_0 (x,t)= \half( {\bf E}^2 + {\bf H}^2 ).
\ee

 For a source-free electromagnetic wave obeying the condition $ {\bf E} \perp {\bf H} $, we can
  choose a special cartesian coordinate system $\{x,y,z\}$, such that
 \be
{\bf E}=(E,0,0),\ \ \ \  {\bf H}=(0,H,0).
 \ee 

 Hence, the zero-order energy velocity in the $z$-direction reads:
\be
w_{0}= -\fr{E\dot{E}+H\dot{H}}{EE'+HH'}
\ee

where the dot and the prime denote the derivative  with respect to time $t$ and direction $z$. respectively.
 Using the Maxwell \eqqs:
 \be
 \dot{H}=c\fr{\pd}{\pd z} E=-cE';\        \ \ \  \dot{E}=-c\fr{\pd}{\pd z} H=-cH';\ \ \    \fr{\pd}{\pd y} E = c\fr{\pd}{\pd x} H = 0; 
 \ee
 we obtain, that 

 \be
w_{0}= c \fr{E H'+H E'}{EE'+HH'}.
\ee
 Moreover, since the electric and magnetic field components are lineary connected through the
"wave impedance" $Z$ by 
\be E=Z H, \ee
 the zero order energy velocity for transversal electromagnetic field propagation is constant:
\be w_{0}= c \fr{Z}{Z^2+1} \ee
and for each real value $Z$ of wave resistance it cannot overcome the light velocity.
 The maximal absolute value $c/2$ can be reached at $Z=\pm 1$.

  Performing the first order (1-PV) energy velocity in a similar way, one obtains

\be w_{I}= -\fr{E'\dot{E}+ H'\dot{H}+E\dot{E}'+ H\dot{H}' }{EE''+HH''+E'^2+H'^2}=c \fr{2Z}{Z^2+1}, \ee
with a maximal value $c$. It can be interpreted in a general meaning as a propagation of a spatial derivative
 of energy density (e.g. a peak of this) and can be called for this reason "force velocity".
  This object can be treated therefore physically in context of interaction transmission or signal transmission.

  An application of this concept to the electrodynamics in media, which could have been of a greatest
  interest in this context, in view of possible application in photonics and near-field optics,
  is actually restricted through the lack of consistent lagrangian formalism in this theory, 
  in the pure field-theoretical sense.  

\section{Conclusions}

The concept of local velocities, applicable for arbitrary fielda, can be applied also for the energy density
as a dynamcal field as well. It implies certain relations between local energy and field velocities. These
relations are in general non-trivial and depends on the construction of the largangian density.  
 For "normal" scalar and free electro-magnatic fields with conventional lagrangians this formalism leads
  also to non-trivial results, which are however apparently meaningful and physically non-controversal,
   and allow a consistent interpretation. 
  It has been argued, for instance to treate the first order energy velocity as a speed of interaction
  in the field. This possibility is of course not uniquie, and the discussion is still open.

\end{document}